\documentclass[aps,pra,twocolumn,amsmath,amssymb]{revtex4}
\usepackage{graphicx} 
\usepackage{bm}
\usepackage{dcolumn}
\usepackage{amsmath}
\usepackage{epstopdf}
\usepackage{natbib}

\begin{document}

\title{Excitonic gap formation and condensation in the bilayer graphene structure}

\author{V. Apinyan\footnote{Corresponding author. Tel.:  +48 71 3954 284; E-mail address: v.apinyan@int.pan.wroc.pl (V. Apinyan).}, T. K. Kope\'{c}}
\affiliation{Institute for Low Temperature and Structure Research, Polish Academy of Sciences\\
PO. Box 1410, 50-950 Wroc\l{}aw 2, Poland \\}

\date{\today}

\begin{abstract}
%
 We have studied the excitonic gap formation in the Bernal Stacked, bilayer graphene (BLG) structures at half-filling. Considering the local Coulomb interaction between the layers, we calculate the excitonic gap parameter and we discuss the role of the interlayer and intralayer Coulomb interactions and the interlayer hopping on the excitonic pair formation in the BLG. Particularly, we predict the origin of excitonic gap formation and condensation, in relation to the farthermost interband optical transition spectrum. The general diagram of excitonic phase transition is given, explaining different interlayer correlation regimes. The temperature dependence of the excitonic gap parameter is shown and the role of the chemical potential, in the BLG, is discussed in details.\end{abstract}

  \maketitle

\renewcommand\thesection{\arabic{section}}

\section{\label{sec:Section_1} Introduction}
%
The bilayer graphene (BLG) structure represents a remarkable interest in modern solid state physics, providing an interesting construction of a material with semiconducting properties \cite{cite_1}. Namely, it is well known that the semiconducting gap of BLG can be tuned by applying the external perpendicular electric field  \cite{cite_2,cite_3, cite_4,cite_5,cite_6,cite_7,cite_8, cite_9,cite_10,cite_11, cite_12, cite_13,cite_14,cite_15}. Namely, a tunable bandgap, up to $20$ meV, has been obtained experimentally \cite{cite_7}, using a bilayer graphene field effect transistor. Meanwhile, an unbiased BLG is a zero-gap semiconductor, characterized by four parabolic bands, where two of them touching each other at zero energy.
 
The excitonic effects in the BLG structures represent another interesting physical phenomenon \cite{cite_16, cite_17,cite_18, cite_19, cite_20, cite_21, cite_22,cite_23,cite_24,cite_25,cite_26}. Recently, the optical response of isolated single- and bilayer intrinsic graphene has been calculated \cite{cite_17}, and the photo-excited states with optical absorption spectra are obtained, using the Gutzwiller-Bethe-Salpeter equation approach.  Particularly, the formation of resonant excitons is shown in the two-dimensional (2D) semi-metallic limit. The first principle calculations, based on the many-electron Green's function approach, have predicted also the existence of bound excitons in one-dimensional (1D) metallic carbon nanotubes \cite{cite_18,cite_19}, which, subsequently, has been verified experimentally \cite{cite_20} using metallic single-walled carbon's nanotubes as a model system. It is demonstrated that the optical transitions, in this 1D metallic systems, are dominated by the excitons with binding energies as large as 50 meV, which significantly exceed that of excitons in most bulk semiconductors \cite{cite_27}.
The binding energy of Wannier-Mott excitons and optical conductivity spectrum of BLG is investigated recently in Ref.\onlinecite{cite_21}. The authors used the simple tight-binding model in the presence of the external gate voltage (extrinsic BLG). Particularly, the effects of excitonic formations on the absorption spectrum are discussed in details, using the Hartree-Fock approximation, and the optical conductance spectrum is calculated, for different values of the external electric field. Another, more recent microscopic study of the optical properties of Bernal-Stacked (BS) BLG structure is given in Ref.\onlinecite{cite_23}, where the influence of the total energy renormalized Coulomb interaction effects on the optical selection rules is determined, and the allowed crossing and non-crossing optical interband transitions are discussed in details. On the other hand, a strong suppression of screening of the interlayer Coulomb interaction, in the case of multiband Cooper pairing, which takes place in a BLG at strong coupling, is discussed in Ref.\onlinecite{cite_28}.

On the other hand, a great experimental and theoretical effort has been dedicated in order to obtain the excitonic condensation in the BLG \cite{cite_29, cite_30, cite_31, cite_32, cite_33, cite_34, cite_35, cite_36, cite_37, cite_38, cite_39, cite_40, cite_41, cite_42, cite_43, cite_44, cite_45}. It is interesting to mention about controversial results, given in Refs.\onlinecite{cite_39, cite_42, cite_43} on the possibilities to obtain the room-temperature excitonic condensate in BLG. Namely, motivated by the large-$N$ limit, and considering the weak-coupling Bardeen-Cooper-Schrieffer (BCS) gap equation, the authors in Refs.\onlinecite{cite_42, cite_43} construct the low energy theory, which gives a negligibly small value of the critical temperature of the excitonic superfluid phase transition. The opposite result is obtained in Ref.\onlinecite{cite_39}, where it has been shown the existence of four independent superfluid orders with the very high transition critical temperatures, when considering the unscreened interlayer coupling interaction.

It appears a natural question, if it is possible to construct a theory, which can unify the obtained previous results for the weak and strong interlayer coupling limits in BLG. 
The present paper gives a detailed recipe how this type of theory could be done. We consider the problem of the excitonic pair formation, in the BLG structures, using the bilayer Hubbard model. By considering the on-site, local interlayer Coulomb interaction, we calculate the excitonic gap parameter in different limits of the interlayer and intralayer Coulomb interactions. We show that, similar to the usual semiconducting systems, an excitonic pairing state is present in the BLG systems, when varying the interlayer Coulomb interaction parameter, from small up to very high values. The principal difference for BLG is that the gap function remains finite up to very large values of the interlayer interaction (this is the limit when the interlayer screening is negligibly small \cite{cite_46}). This is not the case for the intermediate valent semiconductor \cite{cite_47, cite_48, cite_49, cite_50}, or transition metal compounds \cite{cite_51}, where the excitonic insulator state is due to the interband Coulomb interaction. We have calculated the excitonic gap for different values of the interlayer hopping amplitude $\gamma_1$. In accordance with the previous mean-field results, we show that the intralayer repulsive interaction is completely unimportant for the considered problem, and we show that it conducts only to a linear self-consistent (SC) solution of the chemical potential. The obtained results here are related to the excitonic pair formation and condensation in the BLG, and represent a significant interest, when treating different interlayer screening regimes in the BLG system, due to the presence of screening medium between the layers of the BLG structure. Particularly, from our theory, we get both limiting results, discussed above, and we did not found any contradiction between the theories, given in Refs.\onlinecite{cite_39, cite_42, cite_43}.    

The paper is organized as follows: in the Section \ref{sec:Section_2}, we describe the model for treating the BLG system. In the Section \ref{sec:Section_3}, we discuss our theoretical formalism and we obtain the general form of the fermionic action in the intrinsic BLG system at half-filling. In the Section \ref{sec:Section_4}, we derive the self-consistent equations for the excitonic gap parameter and chemical potential. The numerical results are given in the Section \ref{sec:Section_5}.  
In the Section \ref{sec:Section_6}, we discuss our results in touch with the experimental accessibilities, for the BLG structures, and in the Section \ref{sec:Section_7} we will give a short conclusion for our paper.  
%
\section{\label{sec:Section_2} The model}
\begin{figure}
\begin{center}
\includegraphics[width=200px,height=150px]{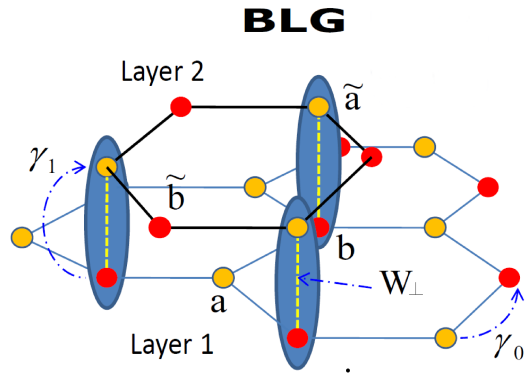}
\caption{\label{fig:Fig_1}(Color online) 
The structure of BS-stacked BLG system. The different sublattice site positions are shown in two different monolayers, of BLG. The excitonic formations are shown in blue structures, between the monolayers of BLG.}
\end{center}
\end{figure}

%
We consider a minimal model for the BLG structure with on-site interlayer interaction. The BLG is composed of two coupled honeycomb layers with sublattices $A$, $B$ and $\tilde{A}$, $\tilde{B}$ placed in the bottom layer and top layer respectively. In the $z$-direction, the layers are arranged according to Bernal Stacking order \cite{cite_34}, i.e. the atoms on the sites $\tilde{A}$ of the top monolayer lie just above the atoms on the sites $B$ of the bottom monolayer graphene, and each monolayer is composed of two interpenetrating triangular lattices (see the BLG structure, given in Fig.~\ref{fig:Fig_1}). We define the bilayer Hubbard Hamiltonian, for the unbiased BLG structure at half-filling, in the form
\begin{eqnarray}
&&H=-\gamma_{0}\sum_{\substack{\left\langle i, j \right\rangle\\ \sigma}}\sum_{l=1,2}\left(X^{\dag}_{li,\sigma}Y_{lj,\sigma}+h.c.\right)
\nonumber\\
&&-\gamma_{1}\sum_{i,\sigma}\left(b^{\dag}_{1i,\sigma}\tilde{a}_{2i,\sigma'}+h.c.\right)-\sum_{i,\sigma}\sum_{l=1,2}{\mu}_{l}n_{li,\sigma}
\nonumber\\
&&+U\sum_{i}\sum_{\substack{\eta=X,Y\\ l=1,2}}\left[\left(n^{\eta}_{li,\uparrow}-1/2\right)\left(n^{\eta}_{li,\downarrow}-1/2\right)-1/4\right]
\nonumber\\
&&+W_{\perp}\sum_{i,\sigma,\sigma'}\left[\left(n^{b}_{1i,\sigma}-1/2\right)\left(n^{\tilde{a}}_{2i,\sigma'}-1/2\right)-1/4\right].
\label{Equation_1}
\end{eqnarray}
Here, we have used the graphite nomenclature \cite{cite_52, cite_53, cite_54} for the hopping amplitudes $\gamma_{0}$ and $\gamma_{1}$. Namely, they corresponds to $\gamma_{0}=t$ (the intraplane hopping amplitude) and $\gamma_{1}=t_{\perp}$ (the interlayer hopping amplitude) in the usual tight-binding notations. The summation $\left\langle i, j \right\rangle$, in the first term in Eq.(\ref{Equation_1}), denotes the sum over the nearest neighbors lattice sites in the separated honeycomb layers. 
The index $l=1,2$ mentions the numbers of single layers in the BLG structure. Particularly, we use $l=1$ for the bottom layer, and $l=2$ for the top layer. The symbol $\sigma$ denotes the spin variables with two possible directions ($\sigma= \uparrow, \downarrow$). The electron operators $X$ and $Y$ in the Hamiltonian are defined in such a way that $X=a$, $Y=b$, for the bottom layer with $l=1$, and $X=\tilde{a}$, $Y=\tilde{b}$, for the top layer with $l=2$. We keep the small letters $a,b$ and $\tilde{a}, \tilde{b}$ for the electrons on the lattice sites $A,B$ and $\tilde{A},\tilde{B}$ respectively, and the notation with tilde is referred to the top layer. Furthermore, $U$, in the Hubbard term in Eq.(\ref{Equation_1}), parametrizes the intralayer Coulomb interaction, and $W_{\perp}$ is the local interlayer Coulomb repulsion. Furthermore, in what follows, we choose $\gamma_{0}=1$, as the unit of energy, and we set $k_{B}=1$, $\hbar=1$.
%
%
Next, $n_{li,\sigma}=n^{X}_{li,\sigma}+n^{Y}_{li,\sigma}$ is the total electron density on site $i$, in the layer $l$, where $n^{X}_{li,\sigma}$ and $n^{Y}_{li,\sigma}$ are electron density operators for different sublattices, defined as $n^{X}_{li,\sigma}=X^{\dag}_{li,\sigma}X_{li,\sigma}$, $n^{Y}_{li,\sigma}=Y^{\dag}_{li,\sigma}Y_{li,\sigma}$.

For a simple treatment, at equilibrium, we suppose the balanced BLG structure with the chemical potentials in both layers that are equal $\mu_{1}=\mu_{2}\equiv \mu$. Thus, we do not suppose the initial hole doping in the bottom layer and we consider the BLG structure with pure electronic layers. We will study the excitonic effects in the BLG system with respect to the half-filling condition assumption in each layer, $\left\langle n_{l} \right\rangle=1$, for $l=1,2$. In order to estimate the energy scales (the excitonic gap, the chemical potential, and the energy bandgaps), related to the excitonic condensation in the BLG, we will discuss in details a particular realistic parameterization for the hopping amplitudes $\gamma_{0}$ and $\gamma_{1}$ (in accordance with Ref.\onlinecite{cite_23}).
%
\section{\label{sec:Section_3} The fermionic action of the BLG system}
\begin{figure}
\begin{center}
\includegraphics[width=250px,height=130px]{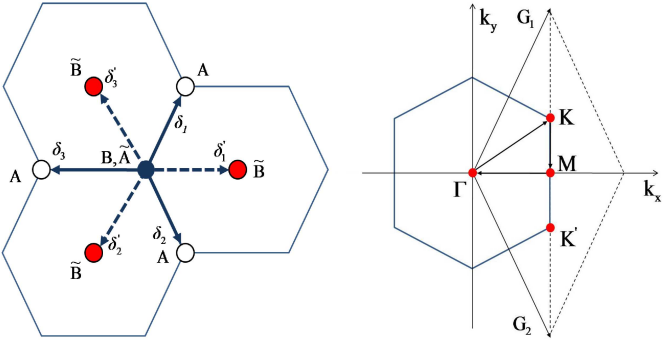}
\caption{\label{fig:Fig_2}(Color online) 
(Left) Top view of the BS-stacked BLG system. The different sublattice site positions are shown (A,B, and $\tilde{A}$, $\tilde{B}$), for two different layers in the BLG, and the nearest neighbors lattice vectors are shown for both layers.(Right) The high symmetry points in the first Brillouin Zone in the honeycomb reciprocal lattice. The basis translation vectors ${\bf{G}}_{1}$ and ${\bf{G}}_{2}$ are shown in the reciprocal space.}
\end{center}
\end{figure}

%
Next, we will pass to the Grassmann's representation for the fermionic variables, and we write the partition function of the system, by employing the imaginary time fermion path integral method \cite{cite_55}.
For this, we introduce imaginary time variables $\tau$, at each lattice site $i$. The time variables $\tau$ vary in the interval $(0,\beta)$, where $\beta=1/T$ with $T$ being the temperature. Then, the grand canonical partition function of the system is 
\begin{eqnarray}
Z=\int\left[D\bar{X}DX\right]\left[D\bar{Y}DY\right]e^{-S\left[\bar{X},X,\bar{Y},Y\right]},
\label{Equation_2}
\end{eqnarray}
where, the action, in the exponent, is expressed as
\begin{eqnarray}
S\left[\bar{X},X,\bar{Y},Y\right]=\sum_{l=1,2}S^{(l)}_{\rm B}\left[\bar{X},X\right]
\nonumber\\
+\sum_{l=1,2}S^{(l)}_{\rm B}\left[\bar{Y},Y\right]+\int^{\beta}_{0}d\tau H\left(\tau\right).
\label{Equation_3}
\end{eqnarray}
The first two terms in Eq.(\ref{Equation_3}), are the fermionic Berry-terms for the layers with indices $l=1,2$. They are given as follows
\begin{eqnarray}
S^{(l)}_{\rm B}\left[\bar{X},X\right]=\sum_{i,\sigma}\int^{\beta}_{0}d\tau \bar{X}_{li,\sigma}(\tau)\frac{\partial}{\partial \tau}X_{li,\sigma}(\tau),
\label{Equation_4}
\newline\\
S^{(l)}_{\rm B}\left[\bar{Y},Y\right]=\sum_{i,\sigma}\int^{\beta}_{0}d\tau \bar{Y}_{li,\sigma}(\tau)\frac{\partial}{\partial \tau}Y_{li,\sigma}(\tau).
\label{Equation_5}
\end{eqnarray}
Here, again, we keep the notations $X_{1i,\sigma}=a_{1i,\sigma}$, $X_{2i,\sigma}=\tilde{a}_{2i,\sigma}$, $Y_{1i,\sigma}=b_{1i,\sigma}$ and $Y_{2i,\sigma}=\tilde{b}_{2i,\sigma}$.
The Hamiltonian of the BLG system, in the last term in Eq.(\ref{Equation_3}), is given in Eq.(\ref{Equation_1}), in the Section \ref{sec:Section_2}, and here we will write $H$ in more convenient form, in terms of the Grassmann's variables $a,b$ and $\tilde{a},\tilde{b}$, corresponding to the layers with $l=1$ and $l=2$,respectively.  
Namely, within the path integral formalism, we have
\begin{eqnarray}
&&H=-\gamma_0\sum_{\substack{\left\langle i,j\right\rangle,\\ \sigma}}\left(a_{1i,\sigma}(\tau)b_{1j,\sigma}(\tau)+h.c.\right)
\nonumber\\
&&-\gamma_0\sum_{\substack{\left\langle i,j\right\rangle,\\ \sigma}}\left(\bar{\tilde{a}}_{2i,\sigma}(\tau)\tilde{b}_{2j,\sigma}(\tau)+h.c.\right)
\nonumber\\
&&-\gamma_1\sum_{i,\sigma}\left(\bar{{b}}_{1i,\sigma}(\tau)\tilde{a}_{2i,\sigma}(\tau)+h.c.\right)
\nonumber\\
&&+U\sum_{\substack{li,\\ \eta=X,Y}}\left[\frac{\left({n^{\eta}_{li}}(\tau)\right)^{2}}{4}-\left({S^{\eta}_{li,z}}(\tau)\right)^{2}\right]
\nonumber\\
&&-\mu_{1}\sum_{i,\sigma}n^{a}_{1i,\sigma}(\tau)-\mu_{2}\sum_{i,\sigma}n^{b}_{1i,\sigma}(\tau)-\mu_{2}\sum_{i,\sigma}n^{\tilde{a}}_{2i,\sigma}(\tau)
\nonumber\\
&&-\mu_{1}\sum_{i,\sigma}n^{\tilde{b}}_{2i,\sigma}(\tau)-W_{\perp}\sum_{i,\sigma,\sigma'}|\chi_{i,\sigma\sigma'}(\tau)|^{2}.
\label{Equation_6}
\end{eqnarray}
We have introduced in Eq.(\ref{Equation_6}) the $z$-component of the generalized spin operator ${\bf{S}}^{\eta}_{li}(\tau)=1/2\sum_{\alpha,\beta = \uparrow, \downarrow}\bar{\eta}_{li,\alpha}(\tau)\hat{\sigma}_{\alpha\beta}\eta_{li,\beta}(\tau)$, for different sublattices, in the layers of BLG. It is defined as $S^{\eta}_{li,z}(\tau)=1/2\left(\eta_{li,\uparrow}(\tau)-\eta_{li,\downarrow}(\tau)\right)$. The chemical potentials $\mu_{1}$ and $\mu_{2}$ are the shifted chemical potentials in the system $\mu_{1}=\mu+U/2$, $\mu_{2}=\mu+U/2+W_{\perp}$. It is noteworthy to mention here that the chemical potentials of electrons on the nonequivalent sublattice sites in the given layer, gets different shifts, in different layers, due to the stacking ordering of the BLG (see in Figs.~\ref{fig:Fig_1} and ~\ref{fig:Fig_2}). The new complex variables $\chi_{i,\sigma\sigma'}(\tau)$ and their complex conjugates $\bar{\chi}_{i,\sigma\sigma'}(\tau)$ are introduced in the last interaction term in Eq.(\ref{Equation_6}), and $\chi_{i,\sigma\sigma'}(\tau)=\bar{b}_{1i,\sigma}\tilde{a}_{2i,\sigma}$. The Hamiltonian, in the form given in Eq.(\ref{Equation_6}), is more suitable for decoupling of four fermionic terms within the Hubbard-Stratanovich type linearisation procedure.
%
%
Furthermore, we perform the real-space linearization of four-fermionic terms in Eq.(\ref{Equation_6}). As an example, we give this procedure for the sublattice-$a$, in the layer-1 of the BLG structure, given in Fig.~\ref{fig:Fig_1}. Namely, we have
\begin{eqnarray}
&&e^{-U/4\sum_{i}\int^{\beta}_{0}d\tau\left(n^{a}_{1i}(\tau)-\frac{2\mu_{1}}{U}\right)^{2}}\sim
\nonumber\\
&&\sim \int\left[DV^{a}_{1}\right]e^{\sum_{i}\int^{\beta}_{0}d\tau\left[-\left(\frac{V^{a}_{1i}(\tau)}{\sqrt{U}}\right)^{2}+iV^{a}_{1i}(\tau)\left(n^{a}_{1i}(\tau)-\frac{2\mu_{1}}{U} \right)\right]}.
\nonumber\\
\label{Equation_7}
\end{eqnarray}
The path integral, in the right hand side (r.h.s.) in Eq.(\ref{Equation_7}), is taken over the decoupling field-variables $V^{a}_{1i}(\tau)$, coupled to the density term. They are introduced at each site position $i$ of the given sublattice and at each time variable $\tau$. The field integral, in r.h.s., in Eq.(\ref{Equation_7}), can be evaluated by the steepest descent method. We get
\begin{eqnarray}
\int\left[DV^{a}_{1}\right]e^{\sum_{i}\int^{\beta}_{0}d\tau\left[-\left(\frac{V^{a}_{1i}(\tau)}{\sqrt{U}}\right)^{2}+iV^{a}_{1i}(\tau)\left(n^{a}_{1i}(\tau)-\frac{2\mu_{1}}{U} \right)\right]}\sim
\nonumber\\
\sim e^{-U/2\sum_{i}\int^{\beta}_{0}d\tau \left(\bar{n}^{a}_{1}-\frac{2\mu_{1}}{U}\right)\left(n^{a}_{1i}(\tau)-\frac{2\mu_{1}}{U} \right)}.
\label{Equation_8}
\end{eqnarray}
Here, in order to obtain the r.h.s., in Eq.(\ref{Equation_8}), we have replaced the field integration over $V^{a}_{1i}$, by using the saddle-point value of the decoupling potential, namely, $\upsilon^{a}_{1}=iU/2\left(\bar{n}^{a}_{1}-2\mu_{1}/U\right)$, and the density average $\bar{n}^{a}_{1}$ is defined with the help of the total action of the system, given in Eq.(\ref{Equation_3}), we have $\bar{n}^{a}_{1}=\left\langle n^{a}_{1i,\uparrow}+n^{a}_{1i,\downarrow}\right\rangle$. The same procedure could be repeated also for the nonlinear density terms, on the other sublattices in the BLG, which contain the density terms $n^{\tilde{a}}_{2i}$ $n^{b}_{1i}$, and $n^{\tilde{b}}_{2i}$. 

The decoupling of the nonlinear density-difference term, in Eq.(\ref{Equation_6}), is also straightforward. Namely, for the $l$-layer and $\eta$-type sublattice variable, we have
\begin{eqnarray}
e^{U\sum_{i}\int^{\beta}_{0}d\tau \left(S^{\eta}_{li,z}(\tau)\right)^{2}}=e^{U/4\sum_{i}\int^{\beta}_{0}d\tau \left(n^{\eta}_{li,\uparrow}-n^{\eta}_{li,\downarrow}\right)^{2}}\sim
\nonumber\\
\sim \int{\left[D\Delta^{\eta}_{c,l}\right]}e^{\sum_{i}\int^{\beta}_{0}d\tau \left[-\left(\frac{\Delta^{\eta}_{c,li}}{\sqrt{U}}\right)^{2}+\Delta^{\eta}_{c,li}\left(n^{\eta}_{li,\uparrow}-n^{\eta}_{li,\downarrow}\right)\right]}.
\label{Equation_9}
\end{eqnarray}
The saddle-point values of the variables $\Delta^{\eta}_{c,li}$ are given by $\delta^{\eta}_{c,l}=U/2\left\langle n^{\eta}_{li,\uparrow}-n^{\eta}_{li,\downarrow}\right\rangle$. Thus, they are proportional to the difference between the average electron densities, with the opposite spin directions. For the simplicity, we suppose the case of the spin-balanced BLG layers, with equal average densities (thus, reflecting the initial, antiferromagnetic ground state of BLG \cite{cite_56})  and for each spin orientation, i.e. $\left\langle n^{\eta}_{li,\uparrow}\right\rangle=\left\langle n^{\eta}_{li,\downarrow}\right\rangle$, and the mean values $\delta^{\eta}_{c,l}$ disappear in the problem. For the case of the half-filling, considered here, we put $\left\langle n^{\eta}_{li,\sigma}\right\rangle=1/2$, for each spin direction $\sigma=\uparrow, \downarrow$, at the equilibrium state . 

Next, we will decouple the last four fermion density term, in Eq.(\ref{Equation_6}).  In this case, we apply the complex form of the Hubbard-Stratanovich transformation \cite{cite_55} for the one-component fermionic-field
\begin{eqnarray}
&&e^{W_{\perp}\sum_{i,\sigma,\sigma'}\int^{\beta}_{0}d\tau|\chi_{i,\sigma\sigma'}(\tau)|^{2}}=
 \nonumber\\
&&=\int{\left[D\bar{\Gamma}D\Gamma\right]}e^{\sum_{i}\int^{\beta}_{0}d\tau -\frac{|\Gamma_{i,\sigma\sigma'}(\tau)|^{2}}{W_{\perp}}}\times
\nonumber\\ 
&&\times e^{\sum_{i}\int^{\beta}_{0}d\tau \bar{\Gamma}_{i,\sigma\sigma'}(\tau)\chi_{i,\sigma\sigma'}(\tau)+\bar{\chi}_{i,\sigma\sigma'}(\tau){\Gamma}_{i,\sigma\sigma'}(\tau)}.
\label{Equation_10}
  \end{eqnarray}
It is not difficult to see that the saddle-point value of the decoupling field variable $\Gamma_{i,\sigma\sigma'}$, introduced in Eq.(\ref{Equation_10}) is related directly to the excitonic gap parameter. Indeed, we have 
  \begin{eqnarray}
  \Delta_{\sigma\sigma'}=W_{\perp}\left\langle \bar{b}_{1i,\sigma}\tilde{a}_{2i,\sigma'}\right\rangle.
  \label{Equation_11}
   \end{eqnarray}
For the next, we will consider the homogeneous BLG structure, when the pairing is with the same orientation of the spin variables, i.e. $\Delta_{\sigma\sigma'}=\Delta_{\sigma}\delta_{\sigma\sigma'}$, excluding the other possibilities of the pairing with the spin inversion (i.e. the spin-flip pairing mechanism \cite{cite_57}). 

Then, we can write the total action of the system in the Fourier representation, given by the transformations
\begin{eqnarray} \eta_{li,\sigma}(\tau)=\frac{1}{\beta{N}}\sum_{{\bf{k}},\nu_{n}}\eta_{{\bf{k}},\sigma}(\nu_{n})e^{i\left({\bf{k}}{\bf{r}}_{i}-\nu_{n}\tau\right)},
\end{eqnarray}
where $\nu_{n}=\pi\left(2n+1\right)/\beta$ with $n=0,\pm1,\pm2,...$, are the Matsubara fermionic frequencies \cite{cite_58}, and $N$ is the total number of sites on the $\eta$-type sublattice, in the layer $l$. We introduce the four-component BLG spinors at each discrete state ${\bf{k}}$ in the reciprocal space and for each spin orientation $\sigma=\uparrow, \downarrow$
 \begin{eqnarray} {\psi}_{{\bf{k}},\sigma}(\nu_{n})=\left[a_{1{\bf{k}},\sigma},b_{1{\bf{k}},\sigma},\tilde{a}_{2{\bf{k}},\sigma},\tilde{b}_{2{\bf{k}},\sigma}\right]^{T}.
\end{eqnarray}
Then, the action of the system, in the reciprocal space, reads as
 \begin{eqnarray} S\left[\bar{\psi},\psi,\bar{\Delta},\Delta\right]=\frac{1}{\beta{N}}\sum_{{\bf{k}},\sigma}\bar{\psi}_{{\bf{k}},\sigma}(\nu_{n})G^{-1}_{{\bf{k}},\sigma}(\nu_{n}){\psi}_{{\bf{k}},\sigma}(\nu_{n}).
  \nonumber\\
  \label{Equation_12}
   \end{eqnarray}
Here, $G^{-1}_{{\bf{k}},\sigma}(\nu_{n})$, is the inverse Green's function matrix, of size $4\times4$. It is defined as
\begin{eqnarray}
\footnotesize
\arraycolsep=0pt
\medmuskip = 0mu
{G}^{-1}_{{\bf{k}},\sigma}\left(\nu_{n}\right)=\left(
\begin{array}{ccccrrrr}
E_{1}(\nu_{n}) & -\tilde{\gamma}_{1{\bf{k}}} & 0 & 0\\
-\tilde{\gamma}^{\ast}_{1{\bf{k}}} &E_{2}(\nu_{n})  & -\gamma_1-\bar{\Delta}_{\sigma} & 0 \\
0 & -\gamma_1-{\Delta}_{\sigma} & E_{2}(\nu_{n}) & -\tilde{\gamma}_{2{\bf{k}}} \\
0 & 0 & -\tilde{\gamma}^{\ast}_{2{\bf{k}}} & E_{1}(\nu_{n}) 
\end{array}
\right).
\label{Equation_13}
\end{eqnarray}

The diagonal elements of the matrix in Eq.(\ref{Equation_13}) are the energy parameters, given by 
\begin{eqnarray}
E_{1}(\nu_{n})=-i\nu_{n}-\mu^{\rm eff}_{1},
\nonumber\\
E_{2}(\nu_{n})=-i\nu_{n}-\mu^{\rm eff}_{2},
\end{eqnarray}
where, the effective chemical potentials $\mu^{\rm eff}_{1}$ and $\mu^{\rm eff}_{2}$, are introduced in the problem 
\begin{eqnarray}
&&\mu^{\rm eff}_{1}=\mu+U/4,
\label{Equation_14}
\newline\\
&&\mu^{\rm eff}_{2}=\mu+U/4+W_{\perp}.
\label{Equation_15}
\end{eqnarray}
The parameters $\tilde{\gamma}_{l{\bf{k}}}$, in Eq.(\ref{Equation_13}), $l=1,2$, are the renormalized (nearest neighbors) hopping amplitudes $\tilde{\gamma}_{l{\bf{k}}}=z\gamma_{l{\bf{k}}}t$, where the ${\bf{k}}$-dependent parameters $\gamma_{1{\bf{k}}}$ and $\gamma_{2{\bf{k}}}$ are the energy dispersions for the BLG layers with $l=1$ and $l=2$, respectively. We have $\gamma_{1{\bf{k}}}=1/z\sum_{{\bf{\delta}}}e^{-i{{\bf{k}}{\bf{\delta}}}}$ (and $\gamma_{2{\bf{k}}}=1/z\sum_{{\bf{\delta}}'}e^{-i{{\bf{k}}{\bf{\delta}}'}}$ for the layer with $l=2$). The parameter $z$, in Eq.(\ref{Equation_13}), is the number of the nearest neighbors lattice sites on the given honeycomb layer, for a given sublattice variable and $z=3$ for each monolayer (see in the left picture in Fig.~\ref{fig:Fig_2}). The nearest-neighbors vectors ${\bf{\delta}}$, in the real space, for the bottom layer-1, (see in the left picture in Fig.~\ref{fig:Fig_2}) are given by ${\bf{\delta_1}}=a/2\left(1,\sqrt{3}\right)$, ${\bf{\delta_2}}=a/2\left(1,-\sqrt{3}\right)$ and ${\bf{\delta_3}}=-a\left(1,0\right)$. For the layer-2, we have abvousely, ${\bf{\delta'_1}}=a\left(1,0\right)$, ${\bf{\delta'_2}}=-a/2\left(1,\sqrt{3}\right)$, and ${\bf{\delta'_3}}=-a/2\left(1,-\sqrt{3}\right)$. Then, for the function $\gamma_{1{\bf{k}}}$, we have
$\gamma_{1{\bf{k}}}=1/3\left(e^{-ik_{x}a}+2e^{i\frac{k_{x}a}{2}}\cos{\frac{\sqrt{3}}{2}k_{y}a}\right)$, where $a$ is the carbon-carbon interatomic distance. By the convention, we put $a\equiv1$, for both layers. For a given geometry, in Fig.~\ref{fig:Fig_2}, it is not difficult to realize that $\gamma_{2{\bf{k}}}=\gamma^{\ast}_{1{\bf{k}}}\equiv\gamma^{\ast}_{{\bf{k}}}$ and, for the renormalized hopping amplitudes, we have $\tilde{\gamma}_{2{\bf{k}}}=\tilde{\gamma}^{\ast}_{1{\bf{k}}}\equiv\tilde{\gamma}^{\ast}_{{\bf{k}}}$, where we have omitted the layer indexes $l$.   

Next, according to the supposition of the spin-balanced BLG structure (remember, that we do not suppose the presence of inhomogeneities or the applied electric field),  we can assume that the pairing gap is real, ($\Delta_\sigma\equiv\Delta=\bar{\Delta}$) and it is not spin-dependent. Therefore, the structure of the matrix is not changing, for the opposite spin directions, i.e. $\hat{G}^{-1}_{{\bf{k}},\uparrow}\left(\nu_{n}\right)\equiv \hat{G}^{-1}_{{\bf{k}},\downarrow}\left(\nu_{n}\right)$. In the next section, we will use the form of the fermionic action, given in Eq.(\ref{Equation_12}), in order to derive the SC equations, which determine the excitonic gap parameter and the chemical potential in the system.  
%
\section{\label{sec:Section_4} The excitonic gap parameter}
%
By basing on the results obtained in the previous section, we will derive here the excitonic gap parameter for the BLG system, under consideration. We use the condition of half-filling, for each layer, in order to find the solution for the chemical potential in the BLG system. For the layer-1, this condition holds, that $\left\langle{n}^{a}_{1}+{n}^{b}_{1}\right\rangle=1$, where ${n}^{a}_{1}$ and ${n}^{b}_{1}$, are the electron densities for $a$ and $b$ type fermions, respectively. The excitonic gap parameter, as it is discussed previously, in the Section \ref{sec:Section_3}, is given as the statistical average $\Delta=W_{\perp}\left\langle \bar{b}_{1i}\tilde{a}_{2i}\right\rangle$, where, just for spin-symmetry's reasons of the matrix, given in Eq.(\ref{Equation_13}), we have restricted to case $\sigma=\uparrow$ and we have omitted the spin indexes for the fermion operators. 

Here, we present only the final results, in the form of the coupled, nonlinear SC equations, for the chemical potential $\mu$ and the excitonic pairing gap parameter $\Delta$. Then, after performing the Matsubara frequency summations, we get 
\begin{eqnarray}
&&\frac{4}{N}\sum_{{\bf{k}}}\sum_{i=1,..,4}\alpha_{i{{\bf{k}}}}n_{\rm F}(\kappa_{i{\bf{k}}})=1,
\label{Equation_16}
\newline\\
&&\Delta=\frac{W_{\perp}(\gamma_1+\Delta)}{N}\sum_{{\bf{k}}}\sum_{i=1,..,4}\beta_{i{{\bf{k}}}}n_{\rm F}(\kappa_{i{\bf{k}}}),
\label{Equation_17}
\end{eqnarray}
where the dimensionless coefficients $\alpha_{i{{\bf{k}}}}$, in Eq.(\ref{Equation_16}) with $i=1,..4$, are given as
\begin{eqnarray}
\footnotesize
\arraycolsep=0pt
\medmuskip = 0mu
\alpha_{i{{\bf{k}}}}
=(-1)^{i+1}
\left\{
\begin{array}{cc}
\displaystyle  & \frac{{\cal{P}}^{(3)}(\kappa_{i{\bf{k}}})}{\left(\kappa_{1{\bf{k}}}-\kappa_{2{\bf{k}}}\right)}\prod^{}_{j=3,4}\frac{1}{\left(\kappa_{i{\bf{k}}}-\kappa_{j{\bf{k}}}\right)},  \ \ \  $if$ \ \ \ i=1,2,
\newline\\
\newline\\
\displaystyle  & \frac{{\cal{P}}^{(3)}(\kappa_{i{\bf{k}}})}{\left(\kappa_{3{\bf{k}}}-\kappa_{4{\bf{k}}}\right)}\prod^{}_{j=1,2}\frac{1}{\left(\kappa_{i{\bf{k}}}-\kappa_{j{\bf{k}}}\right)},  \ \ \  $if$ \ \ \ i=3,4,
\end{array}\right.
\nonumber\\
\label{Equation_18}
\end{eqnarray}
where ${\cal{P}}^{(3)}(\kappa_{i{\bf{k}}})$ is the polynomial of third order,  in $\kappa_{i{\bf{k}}}$, namely%

\begin{eqnarray}	{\cal{P}}^{(3)}(\kappa_{i{\bf{k}}})=\kappa^{3}_{i{\bf{k}}}-a_{1{\bf{k}}}\kappa^{2}_{i{\bf{k}}}+a_{2{\bf{k}}}\kappa_{i{\bf{k}}}-a_{3\bf{k}}
\label{Equation_19}
\end{eqnarray}
with the coefficients $a_{i{\bf{k}}}$, $i=1,...3$, given as 
\begin{eqnarray}
&&a_{1{\bf{k}}}=-2\mu^{\rm eff}_{2}-\mu^{\rm eff}_{1},
\label{Equation_20} 
\newline\\
&&a_{2{\bf{k}}}=\mu^{\rm eff}_{1}\left(\mu^{\rm eff}_{2}+2\mu^{\rm eff}_{1}\right)-\Delta^{2}-|\tilde{\gamma}_{{\bf{k}}}|^{2}, 
\label{Equation_21}
\end{eqnarray}
and
\begin{eqnarray}
&&a_{3{\bf{k}}}=-\mu^{\rm eff}_{1}\left(\mu^{\rm eff}_{2}\right)^{2}+\mu^{\rm eff}_{1}\Delta^{2}+\mu^{\rm eff}_{2}|\tilde{\gamma}_{{\bf{k}}}|^{2}.
\label{Equation_22}
\end{eqnarray}
Next, the coefficients $\beta_{i{{\bf{k}}}}$, in Eq.(\ref{Equation_17}) with $i=1,..4$ are given by the relations
 \begin{eqnarray}
 \footnotesize
\arraycolsep=0pt
\medmuskip = 0mu
\beta_{i{{\bf{k}}}}=(-1)^{i}
\left\{
\begin{array}{cc}
\displaystyle  & \frac{{\cal{P}}^{(2)}(\kappa_{i{\bf{k}}})}{\left(\kappa_{1{\bf{k}}}-\kappa_{2{\bf{k}}}\right)}\prod^{}_{j=3,4}\frac{1}{\left(\kappa_{i{\bf{k}}}-\kappa_{j{\bf{k}}}\right)},  \ \ \  $if$ \ \ \ i=1,2,
\newline\\
\newline\\
\displaystyle  & \frac{{\cal{P}}^{(2)}(\kappa_{i{\bf{k}}})}{\left(\kappa_{3{\bf{k}}}-\kappa_{4{\bf{k}}}\right)}\prod^{}_{j=1,2}\frac{1}{\left(\kappa_{i{\bf{k}}}-\kappa_{j{\bf{k}}}\right)},  \ \ \  $if$ \ \ \ i=3,4,
\end{array}\right.
\nonumber\\
\label{Equation_23}
\end{eqnarray}
where ${\cal{P}}^{(2)}(\kappa_{i{\bf{k}}})$ is given as 
\begin{eqnarray}
{\cal{P}}^{(2)}(\kappa_{i{\bf{k}}})=\left(\kappa_{i{\bf{k}}}+\mu^{\rm eff}_{1}\right)^{2}
\label{Equation_24}
\end{eqnarray}
The function $n_{F}\left(x\right)$, in Eqs.(\ref{Equation_16}) and (\ref{Equation_17}), is the Fermi-Dirac distribution function $n_{F}\left(x\right)=1/\left(e^{\beta{x}}+1\right)$.  
The energy parameters $\kappa_{i{\bf{k}}}$, in Eqs.(\ref{Equation_16}),(\ref{Equation_17}) and (\ref{Equation_18}), (\ref{Equation_23}) with $i=1,...4$, define the band structure of the BLG system, with the excitonic pairing interaction therein. They are given by the following relations
\begin{widetext}

\begin{eqnarray}
\kappa_{1,2{\bf{k}}}=\frac{1}{2}\left[\Delta+\gamma_{1}\pm\sqrt{\left(W_{\perp}-\Delta-\gamma_{1}\right)^{2}+4|\tilde{\gamma}_{{\bf{k}}}|^{2}}\right]-\frac{1}{2}\left(\mu^{\rm eff}_{1}+\mu^{\rm eff}_{2}\right),
\label{Equation_25}
\newline\\
\kappa_{3,4{\bf{k}}}=\frac{1}{2}\left[-\Delta-\gamma_{1}\pm\sqrt{\left(W_{\perp}+\Delta+\gamma_{1}\right)^{2}+4|\tilde{\gamma}_{{\bf{k}}}|^{2}}\right]-\frac{1}{2}\left(\mu^{\rm eff}_{1}+\mu^{\rm eff}_{2}\right).
\label{Equation_26}
\end{eqnarray}
\newline\\
\end{widetext}
The exact numerical solution of Eqs.(\ref{Equation_16})-(\ref{Equation_17}), and the changes in the electronic band structure of the BLG system, in the presence of the excitonic pairing, are discussed in the next section of the present paper.  

\section{\label{sec:Section_5} Numerical results}
%
\subsection{\label{sec:Section_5_1} Gap  parameter and chemical potential}
\begin{figure}
\begin{center}
\includegraphics[width=220px,height=154px]{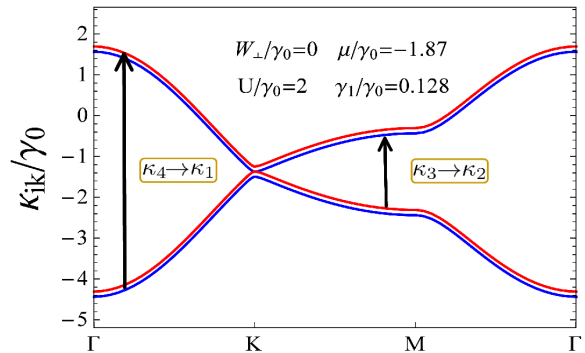}
\caption{\label{fig:Fig_3}(Color online) The electronic band structure of the bilayer graphene in the case of zero interlayer Coulomb interaction and at zero temperature limit $T/\gamma_0=0$. The chemical potential is calculated numerically, after Eqs.(\ref{Equation_16}) and (\ref{Equation_17}).}
\end{center}
\end{figure}

%
Here, we present the numerical results obtained by solving the SC equations for the excitonic pairing gap parameter in the BLG system. First of all, let's mention that the electronic band structure given by the band energies $\kappa_{i{\bf{k}}}$ in Eqs.(\ref{Equation_25})-(\ref{Equation_26}), for the case of the zero pairing gap, is given in Fig.~\ref{fig:Fig_3}. We see that for the particular case of the zero interlayer interaction $W_{\perp}/\gamma_0=0$ and electron-hole pairing $\Delta/\gamma_0=0$, the theory, evaluated here, gives the usual four-band result for the BLG, with two parabolic energy bands $\kappa_{2}$ and $\kappa_{3}$ (for the convenience, hereafter, we will omit the wave vectors near of the band-energy notations), by touching each to the other at the Dirac's points $K$ and $K'$ and corresponding to the $\kappa_3\rightarrow \kappa_2$ optical interband transitions in the BLG system, and two other bands $\kappa_1$ and $\kappa_4$ that are separated by an energy bandgap $E_{g}$, of order $E_g/\gamma_0= 2\gamma_1/\gamma_0=0.256$, in well agreement with the previous results for that case \cite{cite_21, cite_23}. This finite bandgap, for the noninteracting BLG system, corresponds to the non-crossing optical interband transitions $\kappa_4\rightarrow \kappa_1$. Nevertheless, the similarity with the noninteracting BLG band structure, the physics, related to the BLG here, is more complicated. As our recent calculations show \cite{cite_59}, the BLG system is in the weak-coupling BCS like pairing state, even for the negligibly small values of the interlayer Coulomb interaction. This is consistent with the weak-coupling results, given in Refs.\onlinecite{cite_42, cite_43}. The case $W_{\perp}/\gamma_0=0$ will be discussed separately furthermore, in this section, in relation with the bar chemical potential appearing in the system and zero momentum interlayer Fulde–Ferrell–Larkin–Ovchinnikov (FFLO) pairing states at the Dirac's neutrality points.  
%
%

\begin{figure}
\begin{center}
\includegraphics[width=180px,height=360px]{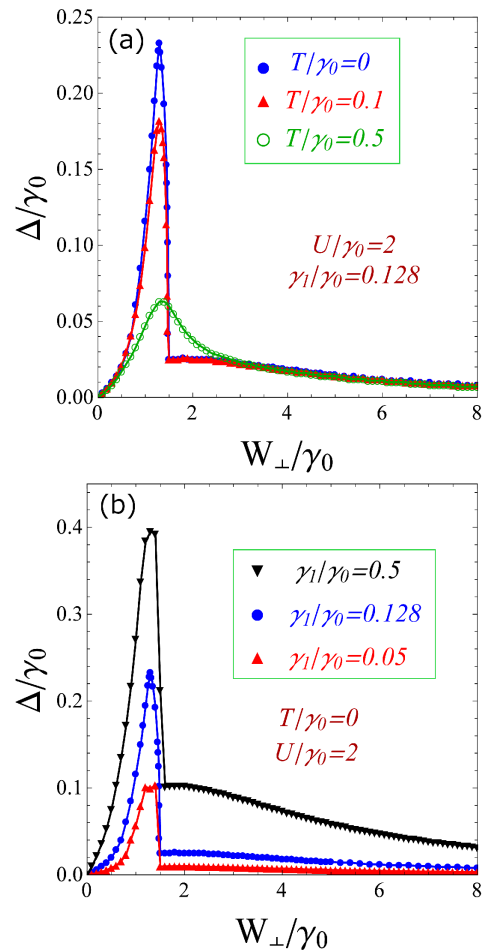}
\caption{\label{fig:Fig_4}(Color online) The excitonic pairing gap parameter $\Delta/\gamma_{0}$, as a function of the interlayer Coulomb interaction $W_{\perp}/\gamma_0$. Different values of temperature (see in the top panel- (a)), and different interlayer hopping amplitudes (see in the bottom panel-(b)) are considered.}
\end{center}
\end{figure}

In Fig.~\ref{fig:Fig_4}, (see in both panels (a) and (b)) we have presented the exact numerical solution for the excitonic pairing gap parameter in the BLG system. The finite-difference approximation method is used in numerical evaluations, which retains the fast convergence of the Newton's algorithm \cite{cite_60}. The convergence of the numerical solution procedure is assumed to be achieved, if all quantities ($\Delta$ and $\mu$) are determined with a relative error of order $10^{-7}$. In the upper panel-(a), in Fig.~\ref{fig:Fig_4}, the solution for the excitonic gap parameter is presented, as a function of the interlayer Coulomb interaction parameter $W_{\perp}/\gamma_0$, for a fixed value of the interlayer hopping amplitude $\gamma_1/\gamma_0=0.128$. A very wide range of the interaction parameter is considered that includes both weak and strong interlayer coupling limits, which, in turn, correspond to different screening regimes in the BLG. The weak interlayer interaction region, in Fig.~\ref{fig:Fig_4}, would correspond to the strong screening effects of the interlayer medium in BLG (after the results in Refs.\onlinecite{cite_42, cite_43, cite_44, cite_45}, this region of the interlayer interaction is very tiny). The higher values of $W_\perp/\gamma_0$ (including the value at which the unscreened gap parameter $\Delta/\gamma_0$ is maximal) correspond to the unscreened interaction regime. The dynamic screening effects are out of the scope of the present paper (see in Ref.\onlinecite{cite_61}, for more details). We see that the behavior of the excitonic pairing gap, versus the interlayer interaction, demonstrates a usual excitonic-insulator type behavior, typical to the case of usual intermediate-valent semiconductor systems \cite{cite_47, cite_48, cite_49}, where the same state is present,      
when considering the variation of the pairing gap parameter as a function of the intralayer interaction parameter $U$. Different values of the temperature are considered in the upper panel-(a), in Fig.~\ref{fig:Fig_4}. The robustness of the excitonic state versus the strong values of the interlayer interaction strength, shown in Fig.~\ref{fig:Fig_4}, is in good agreement with the strong coupling theories, given recently in the Refs.\onlinecite{cite_41, cite_62}. Especially, in the last reference, the strong screening is suggested in the BCS limit of the excitonic phase transition, while in the mixed phase (free excitonic paire+BEC) and pure BEC phase, the screening has been shown as completely unimportant and the presence of the excitonic gap parameter strongly suppresses the anomalous screening polarization function, in that case. 
   
In the lower panel-(b), in Fig.~\ref{fig:Fig_4}, the solution for the excitonic gap parameter is given, for different values of the interlayer hopping amplitude $\gamma_1/\gamma_0$. We see here, that the amplitude of the pairing gap parameter is decreasing more drastically (in comparison with its temperature dependence, given in the upper panel-(a), in Fig.~\ref{fig:Fig_4}), when decreasing the interlayer hopping amplitude $\gamma_1/\gamma_0$. We presume that the behavior of the excitonic gap, obtained in Fig.~\ref{fig:Fig_4}, corresponds well with the $\kappa_4 \rightarrow \kappa_1$ non-crossing absorption spectrum, discussed in Ref.\onlinecite{cite_23}, where it is shown that the absorption spectrum, corresponding to the non-crossing band transitions, vanishes in the near-infrared region, by reflecting a small bandgap in the energy spectrum.  
%
%
\subsection{\label{sec:Section_5_2} The bar chemical potential and charge neutrality point}
\begin{figure}[!ht]
\begin{center}
\includegraphics[width=190px,height=520px]{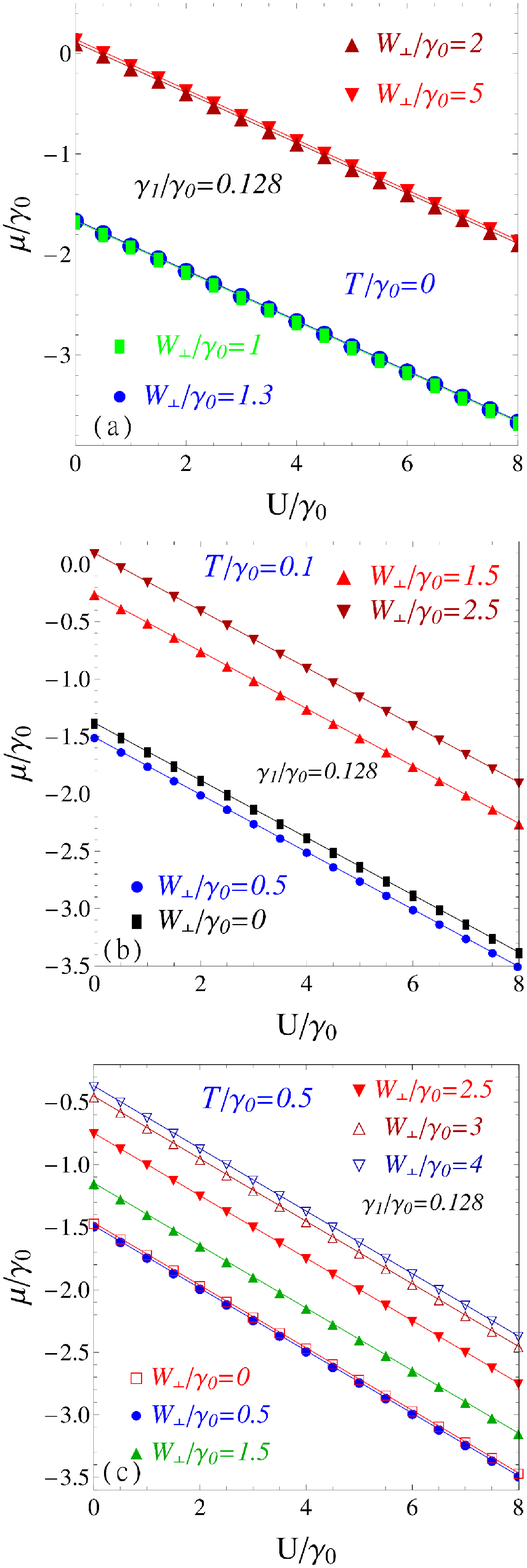}
\caption{\label{fig:Fig_5}(Color online) The solution of the chemical potential, as a function of the intralayer Coulomb interaction parameter $U/\gamma_0$. Different values of temperature and different values of the interlayer Coulomb interaction parameter $W_{\perp}/\gamma_0$ are considered.}
\end{center}
\end{figure}

\begin{figure}[!ht]
\begin{center}
\includegraphics[width=180px,height=180px]{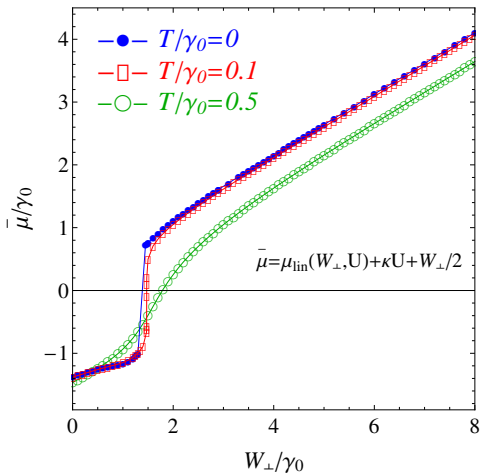}
\caption{\label{fig:Fig_6}(Color online) The effective bar chemical potential in the BLG system. The interlayer hopping amplitude is set at $\gamma_1/\gamma_0=0.128$.}
\end{center}
\end{figure}

%
As the numerical calculations show, the excitonic gap is unchanged when    one varies the intralayer Coulomb interaction parameter $U/\gamma_0$, and only the chemical potential gets modified in that case, as it is presented in Fig.~\ref{fig:Fig_5}. We see, in Fig.~\ref{fig:Fig_5} hereafter, (see in the panels (a)-(c), in Fig.~\ref{fig:Fig_5}) that with the change of the intralayer Coulomb interaction parameter $U/\gamma_0$, the chemical potential $\mu/\gamma_0$, obtained from the system of SC equations, is decreasing linearly, for the fixed value of the parameter $W_{\perp}/\gamma_0$.
%
%
When one changes the values of $W_{\perp}/\gamma_0$, the chemical potential gets the parallel translations in the plane $(\mu/\gamma_0,U/\gamma_0)$. Furthermore, the linear result for the chemical potential solution, as a function of the intralayer $U/\gamma_0$, could be understood after the form of Eqs.(\ref{Equation_14}) and (\ref{Equation_15}), given in the Section \ref{sec:Section_3}. Indeed, if we sum those equations, we get an important equation for the BLG system, namely
\begin{eqnarray}
\mu=-\kappa{U}+\bar{\mu}-\frac{W_\perp}{2},
\label{Equation_27}
\end{eqnarray}
where $\bar{\mu}$ is the bar chemical potential in the BLG system, defined as
\begin{eqnarray}
\bar{\mu}=\frac{\mu^{\rm eff}_{1}+\nu^{\rm eff}_{2}}{2}.
\label{Equation_28}
\end{eqnarray}
It is the average \textit{effective} chemical potential in the system. The form of the chemical potential, in Eq.(\ref{Equation_27}), implements, well, the exact numerical result for $\mu$, shown in Fig.~\ref{fig:Fig_5}. The linear slope of the function, given in Eq.(\ref{Equation_27}) above, is equal to $\kappa=1/4$ and corresponds exactly to the slope of exact $\mu$, in Fig.~\ref{fig:Fig_5}. Let's mention also that after the exact result of $\mu$ in Fig.~\ref{fig:Fig_5}, we are able now to calculate the bar chemical potential $\bar{\mu}$, given in Eq.(\ref{Equation_28}), for a given $W_\perp$ and $U$. The bar chemical potential of BLG, calculated in this way, reflects correctly the particle-hole symmetry in the excitonic pairing region presented in Fig.~\ref{fig:Fig_4}.
In Fig.~\ref{fig:Fig_6}, we have shown the bar chemical potential variation with the interlayer electron-electron interaction parameter. Different values of temperature are considered in Fig.~\ref{fig:Fig_6}. We see that there is a finite jump of the chemical potential at $T/\gamma_0=0$, while for higher temperatures this jump of $\bar{\mu}/\gamma_0$ is smeared. A very similar behavior of the bilayer chemical potential is observed recently, in Ref. \onlinecite{cite_63}, by direct measurement of the chemical potential of BLG as a function of its carrier density. For this purpose, a double-BLG heterostructure has been built, and the bottom bilayer chemical potential has been mapped along the charge neutrality line of the top bilayer. The critical values of the interlayer interaction parameter calculated here $W^{\rm cr}_{\perp}/\gamma_0=1.3$ (or, when $W_{\perp}=3.38$ eV) corresponds to the charge neutrality (i.e. when $n_{B}=0$, where the subscript $B$ indicates the bottom bilayer) of the bottom bilayer in \onlinecite{cite_63}, which could be reached at the back gate voltages of order $V_{BG}\sim -17 $ V, in the Coulomb drag measurements of massless fermions in the BLG \cite{cite_64}. It has been shown that the charge neutrality point for the bottom bilayer is achieved in the electron-electron type bilayer regime, in the Coulomb drag measurements, where the drag resistivity is shown as negative (see in Ref.\cite{cite_64}). Note, that the mentioned back gate potential value, is larger than $W^{\rm cr}_{\perp}/\gamma_0=1.3$ by about one order of magnitude, and induces the same effect for the chemical potential of BLG. On the other hand, it has been indicated in Ref.{\onlinecite{cite_65}} that the interlayer interaction is much weaker in the double BLG constructions compared with the double monolayer graphene and the reason for this is the effect of finite carrier density $n_{T}$ induced in the top BLG, when gating the bottom BLG at $V_{\rm BG}\neq 0$. Consequently, the observed dependence of the BLG's chemical potential on the carrier density, in Ref.{\onlinecite{cite_63}}, is much lower in intensities than the similar effect in our case of a single BLG, when considering the dependence on the electron-electron interaction strength. Note, that the interlayer electron-electron interaction could be modified either, by applying a finite gate voltage to the bottom layer of BLG, or by applying a finite interlayer gate to the top layer. It is important to note that the applied gate to the bottm layer induces the carrier density also in the top layer (see in Ref.\onlinecite{cite_63}, on the example of the double BLG)

\begin{figure}[!ht]
\begin{center}
\includegraphics[width=180px,height=180px]{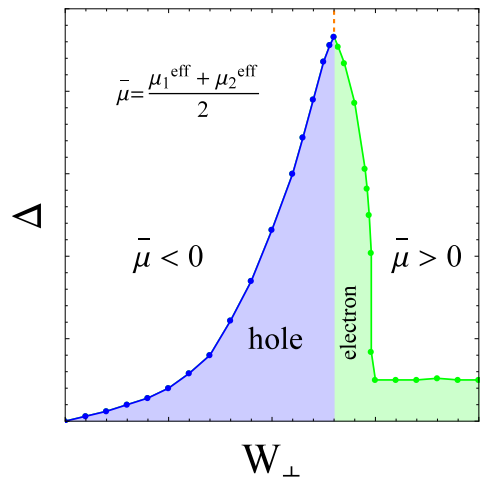}
\caption{\label{fig:Fig_7}(Color online) The phase diagram of the excitonic phase transition in the BLG system. The signs of the bar chemical potential are shown above and below the value $W^{\rm cr}_\perp/\gamma_0=1.3$, at which $\Delta=\Delta^{\rm max}$.}
\end{center}
\end{figure}

Surprisingly, the change of the bar chemical potential's sign, presented in Fig.~\ref{fig:Fig_7}, corresponds well to the similar effect of $\mu$ in Ref.\onlinecite{cite_63}, when passing through the charge neutrality point $n_{B}=0$ of the BLG. Therefore, we can conclude that the theory here for the BLG system could describe the effect of the chemical potential as well as the experiments on the double BLG systems, when considering the charge density variation and the neutrality point. The value $W^{\rm cr}_{\perp}/\gamma_0=1.3$ corresponds to the optimal value of interlayer electron-electron interaction, at which the BLG is unstable with respect to the interlayer exciton formation and condensation. Indeed, if the excitonic gapped state occurs in a system (not especially in BLG), one would expect to see a sharp increase in the resistivity, because, in this case, there are less carriers that occupy the conduction band \cite{cite_66}. This type of behavior has been observed in the narrow-gap semiconductors \cite{cite_67, cite_68}. Turning to the BLG and Ref.\onlinecite{cite_64}, the bottom and top layer resistivities have been measured at the finite, sufficiently high temperatures (in order to escape the mesoscopic coherence effects, which manifest at low temperatures), as a function of layer charge densities. As the results show there, both resistivity curves exhibit sufficiently large maximums at the charge neutrality points in both layers. Thus, we can expect an enhancement of the strong gapped state at this regime. Note that the simultaneous charge neutrality (double neutrality point, NDP) in both layers in BLG has been achieved recently in Ref.\onlinecite{cite_63}, when considering the double BLG heterostructures with the applied interlayer gate at the top bilayer and by considering the BLG resistances and bottom BLG chemical potential along the charge neutrality of the top BLG. The measured bottom bilayer resistance along the charge neutrality of the top BLG shows again large maximum, indicating about the gapped state in the BLG. 
When changing the additional carrier densities in graphene monolayers in BLG, the interlayer electron-electron Coulomb interaction gets strongly modified. \textit{It is very important to mention that the problem of the half-filling BLG, considered here with different interlayer electron-electron interaction regimes, is alternatively equivalent to the problem of gating the initially neutral BLG, and by inducing the variable charge densities in both layers, which, in turn, lead to the modification of the interlayer Coulomb potential, until the next charge equilibrium.} Thus, we have tried to model the system as simple as possible, but, simultaneously, we kept the essential physics therein.
t%
%
Although we have an explicit analytical dependence of $\bar{\mu}$ on the intralayer Coulomb interaction parameter $U$, the numerical values of bar chemical potential, calculated along the linear solutions of the chemical potential, are independent with respect to the variation of $U$, for a given value of the electron-electron interaction parameter $W_{\perp}$. Therefor, we can deduce that the intralayer interaction parameter induces a constant screening, in the BLG, by defining the bar chemical potential, which, furtheremore, is unchanged when varying $U/\gamma_0$. The sign of $\bar{\mu}$, at the vicinity of the critical value $W^{\rm cr}_{\perp}/\gamma_0=1.3$, for $T/\gamma_0=0$, reflects correctly the electron-hole pair formation and condensate region, as it is indicated in Fig.~\ref{fig:Fig_7} here. In addition, we have shown, in Fig.~\ref{fig:Fig_7}, the possible pairing regions: BCS, mixed (free, uncorrelated pairing+BEC), and BEC in the BLG system. 
%
%
\subsection{\label{sec:Section_5_3} The case $W_{\perp}/\gamma_0$: FFLO+BCS limit}
\begin{figure}[!ht]
\begin{center}
\includegraphics[width=240px,height=90px]{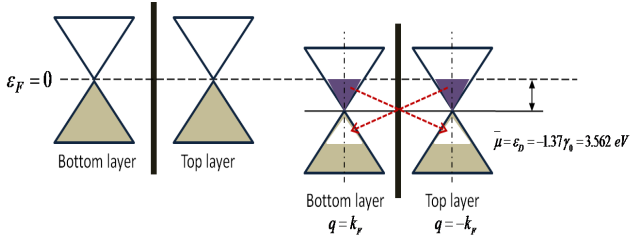}
\caption{\label{fig:Fig_8}(Color online) Band diagram across the noninteracting and unbiased BLG heterostructure. The FFLO pairing state formation with zero center of mass momentum $2q=0$, for the unbiased and noninteracting BLG system.}
\end{center}
\end{figure}

\begin{figure}[!ht]
\begin{center}
\includegraphics[width=160px,height=150px]{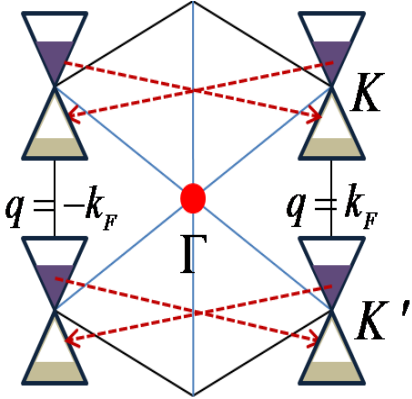}
\caption{\label{fig:Fig_9}(Color online) The excitonic FFLO cross-pairing state formation with zero center of mass momentum, for the single layer graphene.}
\end{center}
\end{figure}

%
Here, we would like to discuss the result presented in Fig.~\ref{fig:Fig_3}, in relation with the bar chemical potential, obtained above. Let's mention that the bands dispersions $\kappa_3$ and $\kappa_2$ are touching each other at the Dirac's points $K$ and $K'$, in the BZ, as it should be for the case of the noninteracting BLG systems  \cite{cite_38}, meanwhile a finite Fermi level solution is given in Fig.~\ref{fig:Fig_3}, at the Dirac's crossing point, coinciding exactly with the bar chemical potential at $W_{\perp}/\gamma_0=0$, presented in Fig.~\ref{fig:Fig_6} (see the value of $\bar{\mu}/\gamma_0$, at $W_{\perp}/\gamma_0=0$, for the case $T/\gamma_0=0$). Thus, the bar chemical potential $\bar{\mu}$ controls the position of the Fermi level in the BLG system. Furthermore, if we calculate the shift of the Fermi energy, caused by the effective bar chemical potential for the noninteracting BLG, we get $\bar{\mu}=-3.562$ eV, which is quite higher than the known results for the undoped neutral BLG \cite{cite_69} (where $\epsilon_{\rm F}\sim \epsilon_{\rm D} \sim -\gamma_1 \sim -0.4 eV$). This fact is related to the presence of strong correlation effect in the BLG, even in the case of absence of the interlayer interactions (unbiased case). Then, by considering the meaningful value for the itralayer hopping amplitude $\gamma_0$, $\gamma_0 = 2.6$ eV \cite{cite_70}, we calculate the Fermi velocity at the Dirac's crossing points in our BLG. Namely, $v_{\rm F}={\bar{\mu}}/{\hbar |k_{\rm F}|}$, where $|k_{\rm F}|$ is the normalized Fermi wave vector at the crossing point (for our case, we have exactly $|k_{\rm F}|=2.418/a$ $A^-1$, and $a$ is the lattice constant in the separated graphene monolayers, we choose $a=2.46$, according to Ref.\onlinecite{cite_70}). We get $v_{\rm F}=0.551\cdot 10^{8} cm/s$, which is nearly 50$\%$ smaller than the Fermi velocity of the gated BLG system \cite{cite_64, cite_71} (with $v_{\rm F}=1.1 \cdot 10^{8}$ cm/s) and about 35 $\%$ smaller than the Fermi energy for the noninteraction in-plane graphene velocity \cite{cite_63, cite_70}. This effect is related to the FFLO electron-hole cross-pairing with a zero momentum at the Dirac's points $K$ and $K'$, between the opposite layers in BLG. For a simple graphene monolayer this type of cross-pairing between the electrons and holes in different layers of is shown in diagrammatically Fig.~\ref{fig:Fig_8}. In Fig.~\ref{fig:Fig_9}, we have shown the similar possible cross-pairing scenario realization in the reciprocal space, for a the simple monolayer graphene sheet.  
%
  
%
%
Another important aspect interpretation of the finite bar chemical potential solution at $W_{\perp}/\gamma_0$ has been given recently in Ref.\onlinecite{cite_59}, where a BCS type excitonic pairing is shown (with no possible supercurrent in the system) at the very weak interlayer Coulomb interaction region, (strong screening), when analyzing the single-particle anomalous momentum distribution functions. The FFLO state formation with zero center of mass momentum, in the unbiased noninteracting BLG structure, is presented in Fig.~\ref{fig:Fig_8}. Another important sign related to the zero momentum FFLO pairing states at $W_{\perp}/\gamma_0=0$ is attributed to the finite and very large particle effective mass in this limit and, as we will show later on in this paper, the effective hole mass is very large for the case $W_{\perp}/\gamma_0=0$. 
%
\subsection{\label{sec:Section_5_4} Condesate transition scenario}

\begin{figure}
\begin{center}
\includegraphics[width=200px,height=400px]{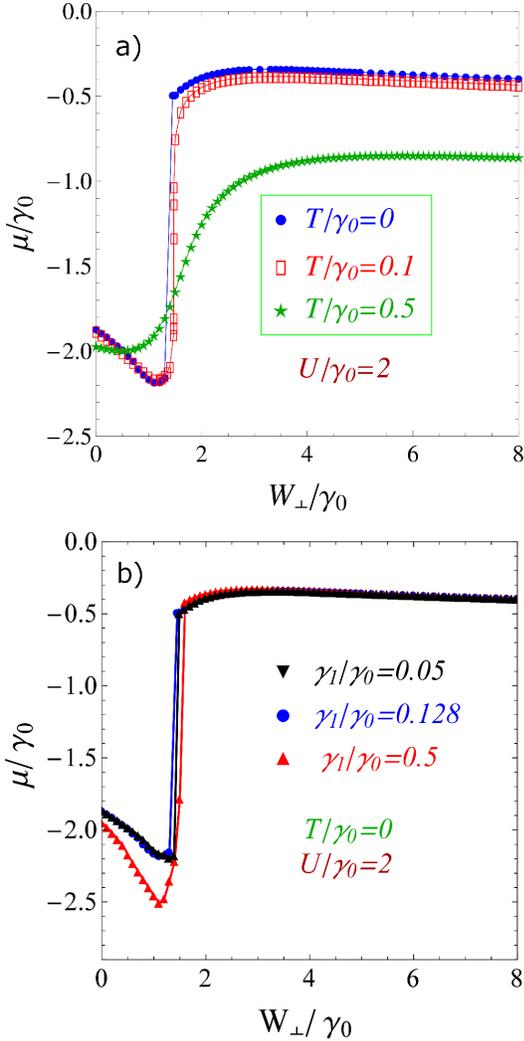}
\caption{\label{fig:Fig_10}(Color online) The solution of the chemical potential as a function of the interlayer Coulomb interaction parameter $W_{\perp}/\gamma_0$. Different values of the  temperature (see in the top panel-(a)) and the interlayer hopping amplitude (see in the bottom panel-(b)) are considered.}
\end{center}
\end{figure}

\begin{figure}
\begin{center}
\includegraphics[width=200px,height=200px]{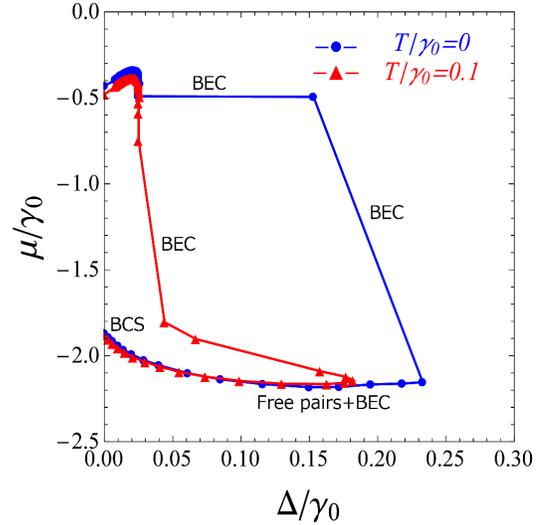}
\caption{\label{fig:Fig_11}(Color online) The chemical potential, as a function of the excitonic gap parameter.}
\end{center}
\end{figure}

In Fig.~\ref{fig:Fig_10}, we have shown the solution for the chemical potential, after the SC equations, given in Eqs.(\ref{Equation_16}) and (\ref{Equation_17}), and discussed in the previous section. In both panels, in Fig.~\ref{fig:Fig_10}, the solution of the chemical potential is presented as a function of the interlayer Coulomb interaction parameter $W_{\perp}/\gamma_0$. In the top panel-(a), in Fig.~\ref{fig:Fig_10}, we have shown also the temperature dependence of the chemical potential, for a fixed value of the interlayer hopping parameter $\gamma_1/\gamma_0=0.128$. In the bottom panel-(b), different values of the parameter $\gamma_1/\gamma_0$ are considered, for the case of zero temperature $T/\gamma_0=0$. We see, in Fig.~\ref{fig:Fig_10}, (see in the top panel-(a), in Fig.~\ref{fig:Fig_10}) that the difference, between lower and upper bounds of the chemical potential, decreases when increasing the temperature. This effect is related to the dependence of temperature of the single-particle excitation gap in the system, defined as 
\begin{eqnarray}
\Delta_{g}=\mu_{\rm max}-\mu_{\rm min}.
\end{eqnarray}
The behavior of $\Delta_g$ as a function of temperature is analog to the temperature dependence of $\Delta/\gamma_0$, given in Fig.~\ref{fig:Fig_4}. The similar temperature dependence of the single-particle excitation gap $\Delta_g$ has been observed for the usual intermediate valent semiconductors discussed recently in Refs.\onlinecite{cite_47, cite_48}.
The function $\Delta_g$ controls, indeed, the pairing interaction in the BLG system, acting as a Clogstone-Zeeman field, similar to the case of superconducting pairing \cite{cite_72}. This is quite analog with the similar effect, given in Ref.\onlinecite{cite_44}, where the  imbalance between the chemical potentials in the layers acts as the splitting field. We see clearly in Fig.~\ref{fig:Fig_10} that there are two extreme upper bounds, for the chemical potential solutions: one corresponds to the limit of the strong screening upper bound $\mu_{ssc}$ ($\mu_{ssc}=-1.87\gamma_0$, i.e $\mu_{ssc}=-4862$ meV, for the case $T/\gamma_0=0$ and for $\gamma_0=2.6$ eV), i.e when the interlayer interaction parameter $W_\perp/\gamma_0 \in (0,0.5)$, while the other upper extreme bound of $\mu_usc$ is situated in the unscreened limit of the interlayer interaction parameter and $\mu_{usc}=-0.34\gamma_0$, i.e $\mu_{usc}=-884$ meV. 
The value of the single-particle excitation gap determines the amplitude and the coherence of the excitonic state at any temperature and at any interlayer hopping (remember that when speaking about the upper and lower bounds of $\mu$ at $T=0$, we should not coincide them with the Fermi levels in different layers, as they are initially supposed the same. These extreme bounds appear when considering the full bandwidth of the interlayer Coulomb interaction $W_\perp/\gamma_0$, as it is presented in Fig.~\ref{fig:Fig_10}). This is in good agreement with the general statements, given in Ref.\onlinecite{cite_61}, concerning of the ``constant gap approximation'', where the wave vector and frequency dependent  self-energy has been calculated and then the center of the wave-vector-energy region has been approximated. 
%
%
This becomes more apparent in Fig.~\ref{fig:Fig_11}, where the solutions of the chemical potential are shown as a function of excitonic gap parameter. We see in Fig.~\ref{fig:Fig_11} that the chemical potential shows a ``hysteresis''-like behavior, as a function of $\Delta/\gamma_0$. The strong screening effects are important here,  along the lower bound of the hysteresis curve (when considering the increasing low values of the excitonic gap). The upper bound corresponds to the unscreened case. We see also that the excitonic gap parameter reaches its maximal unscreened value ($\Delta_{\rm max}/\gamma_0=0.233$, i.e. $\Delta_{\rm max}=0.605$ eV, for the case $T/\gamma_0=0$ and $\Delta_{\rm max}/\gamma_0=0.182$, i.e $\Delta_{\rm max}=0.473$ eV, for the case $T/\gamma_0=0.1$) when the gap $\Delta_g$ collapses ($\Delta_g/\gamma_0=0$). The excitonic pairing and the condensate transition scenarios, given in Figs.~\ref{fig:Fig_4} and ~\ref{fig:Fig_11}, have been verified recently Ref.\onlinecite{cite_59} after a detailed analyses of the anomalous momentum distribution functions and are converging, completely, with the main results, given in Ref.\onlinecite{cite_62}, where the excitonic superfluidity is discussed properly, in two coupled electron-hole armchair-edge graphene nanoribbons, separated by a thin insulating barrier. It is remarkable to note that for the BLG, at half-filling in each monolayer, we have found the excitonic gap parameter nearly in the same order as in the case of bilayer graphene nanoribbons, given in Ref.\onlinecite{cite_62} (where $\Delta \geq 100$ meV). It is worth to mention that the recent results for the excitonic condensation in the BLG \cite{cite_59}, show that for the strongly unscreened regime of the interlayer interaction, the excitonic BEC state separates completely from the free excitonic pairing region, in the form of a perfect condensate nesting in the special triangular pockets, in the reciprocal space, while, in the strongly screened case, the system behaves like weakly coupled BCS state, thus confirming completely the condensate transition scenario, given here, in Figs.~\ref{fig:Fig_7} and ~\ref{fig:Fig_11}.    
%
%
\subsection{\label{sec:Section_5_5} High interlayer interaction: unscreened case}
\begin{figure}
\begin{center}
\includegraphics[width=200px,height=290px]{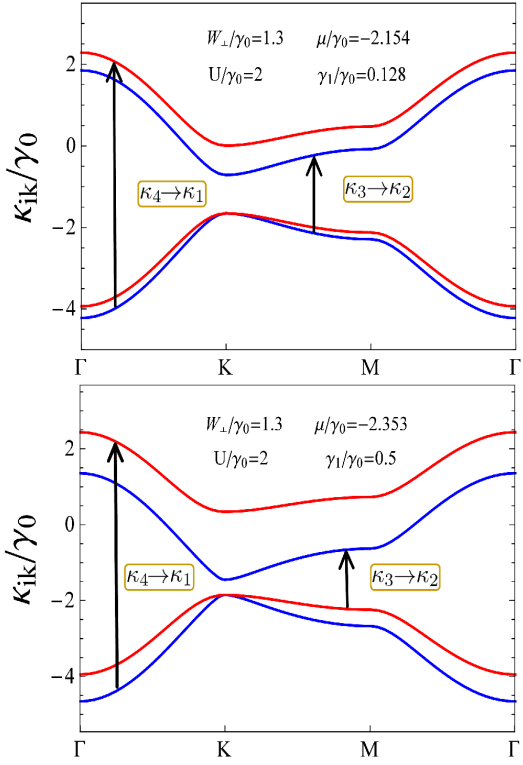}
\caption{\label{fig:Fig_12}(Color online) The electronic band structure of BLG with interacting layers. The interlayer interaction parameter is fixed at $W_{\perp}/\gamma_0=1.3$. The interlayer hopping amplitude is set at $\gamma_1/\gamma_0=0.128$ (see in the top panel) and $\gamma_1/\gamma_0=0.5$ (see in the bottom panel). The zero temperature limit is considered for both cases.}
\end{center}
\end{figure}

\begin{figure}
\begin{center}
\includegraphics[width=200px,height=290px]{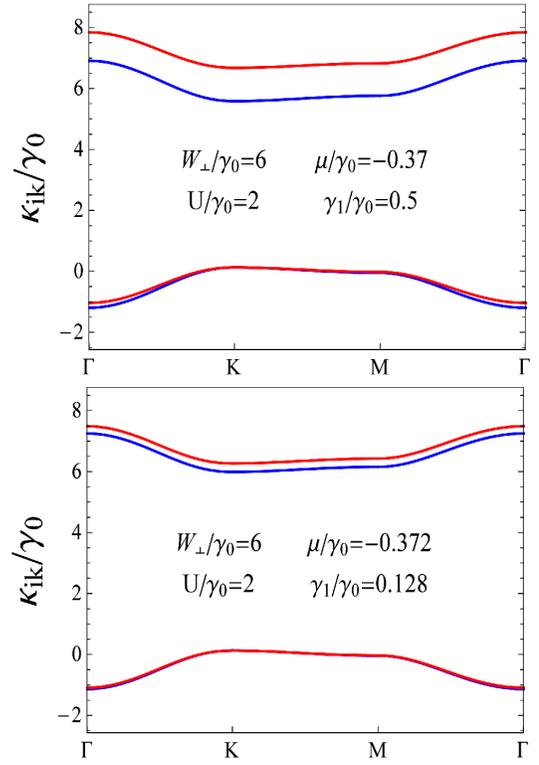}
\caption{\label{fig:Fig_13}(Color online) The electronic band structure of BLG with the interacting layers, for the high interlayer interaction parameter, fixed at $W_{\perp}/\gamma_0=6.0$. The interlayer hopping amplitude is set at $\gamma_1/\gamma_0=0.128$ (see in the top panel) and $\gamma_1/\gamma_0=0.5$ (see in the bottom panel). The zero temperature limit is considered for both cases.}
\end{center}
\end{figure}

%
In Figs.~\ref{fig:Fig_12} and ~\ref{fig:Fig_13}, the band structure of the BLG system is shown, for the case of finite interlayer Coulomb interaction. In Fig.~\ref{fig:Fig_12}, the interlayer interaction parameter is fixed at the value $W_{\perp}/\gamma_0=1.3$, which corresponds to the maximum value of the excitonic pairing gap parameter (see in Fig.~\ref{fig:Fig_4}, where $\Delta_{\rm max}/\gamma_0=0.233$ at $W_{\perp}/\gamma_0=1.3$ and for $T/\gamma_0=0$). We
%
%
We see that, due to the excitonic effects in the BLG system, there is a finite bandgap $E_{g}$ in the electronic band structure of BLG, even for the non-crossing interband transitions $\kappa_3\rightarrow\kappa_2$. The band structure, illustrated in Fig.~\ref{fig:Fig_12}, shows the band structure modifications (in comparison with the zero interaction limit, given in Fig.~\ref{fig:Fig_3}), due the interlayer interaction, and the excitonic pair formations. In this case, a sufficiently large bandgap is opening at the Dirac's points, for the non-crossing subbands $\kappa_3 \rightarrow \kappa_2$ of order $E_{g}/\gamma_0=0.939$ (at the place of $E_{g}/\gamma_0=0.256$, for the noninteracting case). The other, $\kappa_4 \rightarrow \kappa_1$ non-crossing interband transition bandgap is obtained as $E_{g}/\gamma_0=1.661$. Furthermore, the energy bandgap is very sensible also to the changes of the interlayer hopping parameter $\gamma_1/\gamma_0$. We see, in the bottom panel, in Fig.~\ref{fig:Fig_12} that, when increasing the parameter $\gamma_1/\gamma_0$, the non-crossing energy gap $E_g/\gamma_0$ for the interband transitions $\kappa_4 \rightarrow \kappa_1$ is increasing, and for $\gamma_1/\gamma_0=0.5$, we have $E_{g}/\gamma_0=2.195$, while the bandgap, for the transitions $\kappa_3 \rightarrow \kappa_2$, is decreasing and $E_{g}/\gamma_0=0.405$. Thus, we have shown in Figs.~\ref{fig:Fig_13} that the interlayer hopping amplitude has an opposite effect on the energy bandgaps, corresponding to the non-crossing optical transition $\kappa_3 \rightarrow \kappa_2$ and $\kappa_4 \rightarrow \kappa_1$. In Fig.~\ref{fig:Fig_13}, we have considered the electronic band structure in the strong interlayer interaction limit $W_{\perp}/\gamma_0=6$. Two different values of the hopping amplitude $\gamma_1/\gamma_0$ are considered. 
%
%
For the case $\gamma_1/\gamma_0=0.128$, (see in the top panel, in Fig.~\ref{fig:Fig_13}), we get for the $\kappa_4 \rightarrow \kappa_1$ interband transitions $E_{g}/\gamma_0=6.14$, and $E_{g}/\gamma_0=5.86$, for the $\kappa_3 \rightarrow \kappa_2$ interband transitions. For the higher value of the interlayer hopping amplitude $\gamma_1/\gamma_0=0.5$ (see in the bottom panel, in Fig.~\ref{fig:Fig_13}), we get $E_{g}/\gamma_0=6.547$, for the transitions $\kappa_4 \rightarrow \kappa_1$, and $E_{g}/\gamma_0=5.453$, for the transitions $\kappa_3 \rightarrow \kappa_2$. We observe that for the strong values of the interlayer coupling interaction, the general behavior of the bandgap is the same, as in the previous case of the weak interlayer coupling $W_{\perp}/\gamma_0=1.3$. The difference is that for the strong interaction limit, the band structure branches, corresponding to the inter-subband transitions $\kappa_4\rightarrow\kappa_3$ and $\kappa_2\rightarrow\kappa_1$, become close and parallel to each other, overall the high symmetry points, in the FBZ, and the changes in the non-crossing energy gaps are not significant, when changing the interlayer hopping amplitude $\gamma_1/\gamma_0$, i.e. the system becomes more stable with respect to the interlayer hopping. From the discussion above, and taking into account the behavior of the excitonic pairing gap parameter $\Delta/\gamma_0$, when changing the interlayer hopping parameter (see in the upper panel-(a), in Fig.~\ref{fig:Fig_4}), we can suppose that the pairing gap $\Delta/\gamma_0$ corresponds well to the non-crossing interband transitions $\kappa_4\rightarrow\kappa_1$, for both, weak and strong interlayer coupling limits.           
%
\section{\label{sec:Section_6} Discussion}
%
We will discuss shortly the effects of the excitonic pairing on the non-crossing optical transitions in the BLG system, and we will estimate the energy scales related to the excitonic pairing gap, chemical potential \cite{cite_73}, and the bandgaps. 
Turning to the results, given in the Section \ref{sec:Section_5}, let's consider a concrete realistic value for the nearest neighbors hopping amplitude $\gamma_0$ in the layers. According to the work, in Ref.\onlinecite{cite_23}, we will consider the tight-binding value for intralayer hopping amplitude $\gamma_0=2.6$ eV. For this case, we have $\gamma_1=0.128\gamma_0=0.33$ eV, corresponding to the plot of the energy bands, presented in Fig.~\ref{fig:Fig_3}, where we have supposed the zero pairing interaction in the BLG system, with the zero pairing gap parameter $\Delta/\gamma_0=0$, as it follows from the exact numerical results, presented in Fig.~\ref{fig:Fig_4}, in the Section \ref{sec:Section_5}. There is a finite gap, in Fig.~\ref{fig:Fig_3}, between the non-crossing subbands $\kappa_4$ and $\kappa_1$, of order $2\gamma_1=0.256\gamma_0=0.665$ eV, corresponding to the $\kappa_4\rightarrow\kappa_1$ interband optical transitions. This value is very close to the value, given in Ref.\onlinecite{cite_23}. The intensity in the absorption spectrum corresponding to the energetically higher transition $\kappa_4\rightarrow\kappa_1$, is lower, as it has been discussed in Ref.\onlinecite{cite_23}, due to the strong overlap of the low-energy edges of the $\kappa_4\rightarrow\kappa_1$ interband transition. Additionally, there is no pronounced absorption peak in the near infrared region, but only a small bandgap appears, of order $2\gamma_1$, in that region of the photon energy spectrum. 

For a finite value of the interlayer Coulomb interaction parameter $W_{\perp}=1.3\gamma_0=3.38$ eV, we calculate the excitonic pairing gap $\Delta$, chemical potential $\mu$ and excitonic bandgap $E_g$ at the Dirac's points $K$, $K'$. These important physical parameters in the system, calculated in the frames of our theoretical model, are given in Table~\ref{tab:a} below, for the case $T/\gamma_0=0$. We don't discuss here the crossing bandgaps, corresponding to the crossing interband transitions $\kappa_3\rightarrow\kappa_1$ and $\kappa_4\rightarrow\kappa_2$, because they don't contribute to the absorption spectrum, due to the vanishing of corresponding optical matrix elements, away from the Dirac's nodal points (see the discussion on that subject, given in Ref.\onlinecite{cite_23}). We see that, for the case of the finite interlayer interaction parameter, corresponding to the maximum of the excitonic gap $\Delta=1.3\gamma_0=3.38$ eV (see in Fig.~\ref{fig:Fig_4}), and for the high value of the interlayer hopping amplitude $\gamma_1=0.5\gamma_0=1.3$ eV, the difference, between two values of the bandgap, is of order $\sim 4.65$ eV, which is quite larger, in comparison with the value $\sim 1.87$ eV, obtained for the smaller value of the hopping $\gamma_1=0.128\gamma_0=0.33$ eV. This observation is in good agreement with the results given in Ref.\onlinecite{cite_21}.         
\begin{table}[h!]
  \centering
  \begin{tabular}{{c|c|c}}
    BLG's parameters & ${\gamma_1}=0.33 \ eV$ & ${\gamma_1}=1.3 \ eV$\\
    \hline
    \\
    $\Delta$ (eV) & 0.605 & 1.027\\
    $\mu$ (eV) & -5.6 & -6.11\\
    $E_g\left(\kappa_3\rightarrow\kappa_2\right)$ (eV) & 2.44 & 1.053 \\
    $E_g\left(\kappa_4\rightarrow\kappa_1\right)$ (eV) & 4.31 & 5.707 \\
    
  \end{tabular}
    \caption{The values of the important physical parameters, calculated for the BLG system. The interlayer Coulomb interaction parameter is fixed at $W_{\perp}/\gamma_0=1.3$, and the zero temperature limit is considered.    }
 \label{tab:a}
\end{table}
The excitonic pair formation, discussed here, survives for the values of the interlayer Coulomb interaction, given over the full bandwidth, and can persist up to a very high values of temperature (see the temperature dependence of the excitonic gap parameter, given in the upper panel-(a) in Fig.~\ref{fig:Fig_4}, in the Section \ref{sec:Section_5}). 

We would like also to discuss here the displacement of the Fermi level when varying the interlayer Coulomb interaction parameter. It is well known that the BLG system keeps the Fermi liquid properties when including the self-energy renormalization effects on the quasiparticle spectrum, caused by the many-body interactions in the system (see also the discussion, in Ref.\onlinecite{cite_74}).
Thus, at $T/\gamma_0=0$, the chemical potential is a good approximation for the Fermi level. We can estimate now the Fermi energy, for a given value of the interlayer Coulomb interaction parameter and the effective particle mass, at the large momentum $|\vec{k}|$. The fact that in the BLG we have the massive fermionic particles is related to the finite interlayer hopping amplitude $\gamma_1$, 
which is the energy, needed for the particle transitions $b\rightarrow\tilde{a}$, from the bottom to top layer in BLG. It is known that, at the low energy bands and at the large momentum, the quasiparticle energy spectrum in the BLG system can be interpolated, approximately, to the linear dispersion, \cite{cite_75}, as in the case of the monolayer graphene, i.e. $\epsilon\sim v_{F}|\vec{k}|$, where $v_{F}$ is the Fermi velocity in the monolayer graphene sheets, in the BLG. The recent measurements of the Fermi energy in graphene, using 
a double-layer heterostructure \cite{cite_76}, suggests that we have $v_{F}=1.15\times 10^{8}$ cm/s. In this limit, the effective particle mass \cite{cite_75, cite_76} is given by: $m^{\ast}={k_{F}}/{v_{F}}$ (here, and in the previous formula, we keep the convention, where $\hbar=1$). Let's now consider the case of the zero interlayer interaction. By putting the realistic value for the intralayer hopping amplitude $\gamma_0=3.43 \pm 0.01$ eV \cite{cite_71, cite_77}, we get within our theory $\epsilon_{F}=\mu_{|T=0}=-6.41$ eV, and the Fermi level lies at the vicinity of the upper edge of the valence band, i.e. below of Dirac's crossing energy (see in Fig.~\ref{fig:Fig_3}, in the previous Section \ref{sec:Section_5}). We get for the effective mass $m^{\ast}=-0.851 m_{\rm el}$, where 
$m_{\rm el}$ is the free electron mass. Similarly, we can calculate the effective mass for other values of the interlayer Coulomb interaction parameter $W_{\perp}/\gamma_0$. In general, when calculating the effective mass, at the large momentum, we can use the following expression for the effective mass $m^{\ast}=0.132a\gamma_0m_{\rm el}$, where $a$, is the numerical solution for the chemical potential (see, for details, in Fig.~\ref{fig:Fig_7} in the Section \ref{sec:Section_5}). In Table~\ref{tab:b}, we present the numerical results for the coefficient $a$, and for the renormalized effective mass $m^{\ast}/m_{\rm el}$. A set of the specific values \cite{cite_59} of the interlayer Coulomb interaction parameter $W/\gamma_0$ is considered. 

\begin{figure}
\begin{center}
\includegraphics[width=160px,height=160px]{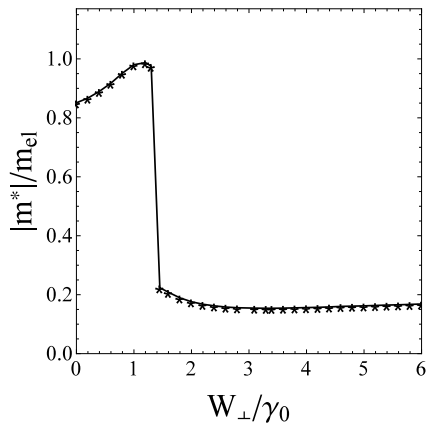}
\caption{\label{fig:Fig_14}(Color online) The effective hole mass in the BLG, in the large momentum limit, as a function of the interlayer Coulomb interaction parameter $W_{\perp}/\gamma_0$. The interlayer hopping amplitude $\gamma_1$ is set at $\gamma_1/\gamma_0=0.128$, and the zero temperature limit is considered.} 
\end{center}
\end{figure}

\begin{table}[h!]
  \centering
  \begin{tabular}{{c|c|c|c|c|c|c|c|c}}
    $W_{\perp}/\gamma_0$ & 0 & 0.8 & 1.2 & 1.3 \\
    \hline
		 \\
    $a$ (eV) & -1.87 & -2.102 & -2.181 & -2.154 \\
    $m^{\ast}/m_{\rm el}$ & -0.851 & -0.952 & -0.987 & -0.975 \\
  \end{tabular}
 \begin{tabular}{{c|c|c|c|c|c|c|c|c}}
  \hline
  \newline\\
   \hline
	$W_{\perp}/\gamma_0$ & 1.45 & 2 & 4 & 6 \\
    \hline
		 \\
    $a$ (eV) & -0.494 & -0.389 & -0.345 & -0.372\\
    $m^{\ast}/m_{\rm el}$ & -0.223 & -0.176 & -0.156 & -0.168\\
  \end{tabular}
    \caption{The effective hole mass in the BLG system, calculated theoretically, for different values of the interlayer Coulomb interaction parameter $W_{\perp}/\gamma_0$. The zero temperature limit is considered.}
 \label{tab:b}
\end{table}
In Fig.~\ref{fig:Fig_14}, we have plotted the variation of the effective hole mass in the BLG, as a function of the local interlayer Coulomb interaction parameter $W_{\perp}/\gamma_0$, for the zero temperature case. The large momentum approximation is considered, and the spectrum is nearly linear, as in the case of the single layer graphene. The non-monotonic behavior of the effective mass with respect to the interaction parameter is related to the behavior of the chemical potential $\mu$ in that case (see the SC solution of the chemical potential in the BLG system, given in Fig.~\ref{fig:Fig_10}, in the Section \ref{sec:Section_5}). Note, that the hole effective mass riches its unscreened value $m^{\ast}=m_{\rm el}$ at $W_{\perp}/\gamma_0=1.3$, i.e. at the value of $W_{\perp}$, corresponding to the double charge neutrality point. A very similar dependence of the effective mass on the bottom BLG's charge density is given in Ref.\onlinecite{cite_63}. The very large effective mass, in the case of the noninteracting BLG. i.e. when $W_{\perp}/\gamma_0=0$, suggests that in this limit there are bound electron-hole pairs in the BLG system, as it is suggested in the Section \ref{sec:Section_5}. In fact, in this limit, the BLG system is in the weak correlated BCS regime \cite{cite_59}, and also the zero momentum FFLO cross-pairing is present in the BLG (see the discussion in the Section{\ref{sec:Section_5}}, and in Figs.~\ref{fig:Fig_8} and ~\ref{fig:Fig_9}).    
%
%

As far as the results in Ref.\onlinecite{cite_77} suggest, the interlayer hopping parameters $\gamma_3$ and $\gamma_4$ (according to the notation of the model of Slonczewski-Weiss-McClure \cite{cite_78}), are unimportant for treating the excitonic properties in the BLG. The first parameter $\gamma_3$, describing the hopping $a-\tilde{b}$, leads to an effective trigonal warping, and for the BLG, this effect is strong, only at the low-energy part of the spectrum. The parameter $\gamma_4$, in turn, describes the interlayer hopping between the lattice sites $a-\tilde{a}$ or $b-\tilde{b}$. This parameter has no influence on the intensity patterns (see the discussion in Ref.\onlinecite{cite_48}). In contrast, the substrate-induced asymmetry, and the interlayer asymmetry (obtained by the anisotropy of the constant energy-maps) could alter the experimental interference patterns and, furthermore, could make an important modification to the single-particle spectral properties. The asymmetry inclusion in the considered problem is out of the subject of the present work. 

Another important experimental aspect, that should be mentioned here, is the influence of the interlayer medium on the interlayer-exciton formation and, especially, the screening effect of it, on the interlayer Coulomb interaction parameter $W_{\perp}/\gamma_0$. For this, the insulating dielectrics have been largely applied, which are thinned to the point, where charge build up and crosstalk adversely affect the performance of the electronic devices. On the other hand, the direct experimental measurement of the excitonic gap parameter is extremely difficult, due to the very short lifetime of excitonic quasiparticles and fast electron-hole recombination effects. In this sense, a replacement of the medium substrate (between the layers, in the BLG) from insulator (such as the porous SiO$_{2}$, or carbon doped SiO$_{2}$) by a doped semiconductor (playing the role of the excitonic bath, and then by stabilizing the states with excitonic quasiparticles), could help to improve experimental measurements on the excitonic effects. Especially, such a semiconducting medium will provide additional donor trap levels in two different layers of the BLG and will improve the excitonic effects considerably. In this case, the BLG will probably exhibit photoluminescence. Specifically, for the intrinsic BLG, such in our treatment, the bandgap, tuned by the interlayer interaction (which could be controlled by varying the interlayer medium transparency, for example), would allow, even for the unbiased BLG, for logic and optoelectronic applications. At the end, we would like to emphasize on the behavior of the hole effective mass in the BLG, as a function of the interlayer interaction parameter. Although
 the half-filling condition in each layer of BLG,  the $W_{\perp}/\gamma_0$-dependence of the effective mass is a direct consequence of the Fermi level SC solutions at $T/\gamma_0=0$ (see the SC solutions of the chemical potential at $T/\gamma_0=0$, in Fig.~\ref{fig:Fig_8}, in the Section \ref{sec:Section_5}).    
          
%
\section{\label{sec:Section_7} Concluding remarks}
%
Summarizing the obtained results, we would like to emphasize on the principal achievements in the present paper. We have studied the problem of the exciton formation in the BLG systems. The main accent of the presented theory is put on the effect of the interlayer Coulomb interaction, which could be fully controlled by switching on and off the interlayer screening by applying the gate voltage to the BLG structure. The excitonic gap parameter, chemical potential and bar chemical potential of the BLG system have been calculated numerically, as a function of the interlayer interaction potential. Particularly, at $T/\gamma_0=0$, the Fermi energy solutions have been obtained, corresponding to different interlayer interaction regimes in the BLG structure, by supposing that the Fermi liquid picture is valid for the BLG. It is remarkable to note, that in the limits of weak and strong interlayer interaction potential the theory, evaluated here, permits to obtain the previous results \cite{cite_39, cite_42, cite_43} on the same subject, and in this sense it is more general. 

Meanwhile, we have reconstructed the interacting band energy curves of the BLG with the excitonic pairing interaction. We have shown that a significant bandgap appears in the energy spectrum when switching the interlayer Coulomb interaction, and we have examined the variation of the band energy curves corresponding to different optical transitions in the system, by varying the interlayer interaction parameter and the temperature. Namely, we have shown that the excitonic pairing and condensation is significant especially in the vicinity of the absorption edges corresponding to the farthermost interband optical transition boundaries in the band energy distribution spectrum. 
Similar to the usual semiconducting or rare-earth compounds, we have obtained an excitonic insulator region, driven by the interlayer Coulomb interaction parameter in the BLG system. In difference with the mentioned materials, this state in the BLG persists up to very high values of the interaction parameter and very high temperatures and is well pronounced in the narrow region of the interlayer interaction straight $W_{\perp}/\gamma_0\in(1,1.4)$. The theory evaluated here permits to clearly distinguish the effects of three different interlayer interaction regimes: weak-coupling BCS regime+FFLO cross-pairing, where the excitonic gap parameter is negligibly small due to the strong screening effects \cite{cite_40, cite_42, cite_43}, the mixed state consisting of the free excitonic pairs + the BEC of excitons, and the excitonic BEC state corresponding the robustness of the excitonic gap parameter (the unscreened case \cite{cite_41, cite_44, cite_45, cite_62}). The principal distinguishing feature of our work is that it is not addressed to the specific limit of the interlayer correlations and we treat the excitonic effects in the BLG as general as possible. The further calculations of the momentum distribution functions and the excitonic density of states in the BLG, within the same theoretical approach \cite{cite_59}, have confirmed of the existance of the phase diagram, presented in Fig.~\ref{fig:Fig_7}. As far, as the results, shown here, the theory strongly suggests 
the possibility of the excitonic condensate states, even at room temperatures.            

We hope that our results will form a solid background to examine furthermore the excitonic effects in the BLG structures. Especially, the context of the excitonic condensation in the intrinsic bilayer structures represents actually a hot research topic for the future. In our opinion, the further analysis of the excitonic density of states and spectral properties will, undoubtedly, confirm the possibility of the excitonic condensation phenomenon in the BLG systems. In the wide prospect, of quantum information and quantum computation, it would be essential to apply the obtained results here, for the double BLG system \cite{cite_79, cite_80} and to examine the biexciton formations in that case and since the exciton-exciton interactions can be manipulated in such double BLG, in order to produce an accurate and efficient degree of control for quantum logic.

%

%
\end{document}